\documentclass[aps,showpacs]{revtex4}
\def\be{\begin{equation}}
\def\ee{\end{equation}}
\usepackage[dvips]{graphicx}

\begin{document}
\draft
%
%

\title{The Coester Line in Relativistic Mean Field Nuclear Matter}
\vskip .3cm
\author{A. Delfino$^{\star}$ ,
M. Malheiro$^{\star}$, 
V. S. Tim\'oteo$^{\star\star}$ , and 
J. S. S\'a Martins$^{\star}$}

\affiliation{$^{\star}$Instituto de F\'\i sica, Universidade Federal Fluminense, \\ 
24210-340, Niter\'oi, R J, Brasil \\ 
$^{\star\star}$Centro Superior de Educa\c c\~ao Tecnológica, \\ 
Universidade Estadual de Campinas, \\ 
13484-370, Limeira, SP, Brasil }

\begin{abstract} 

The Walecka model contains essentially two parameters that are associated
with the Lorentz scalar ( $S$ ) and vector ( $V$ ) 
 interactions. These parameters are related to a two-body interaction
  consisting of $\,S\,$ and $\,V\,$, imposing the
 condition that the two-body binding energy is fixed. 
 We have obtained a set of different values for
 the nuclear matter binding energies ( $B_{N}$ ) at
  equilibrium densities ( $\rho_{o}$ ). We investigated the existence of a
linear correlation between $B_{N}$ and $\rho_{o}$, claimed to be
 universal for nonrelativistic systems and usually known 
 as the Coester line, and found an approximate linear correlation only
if  $\,V\,-\,S\,$ remains constant. It is shown that the relativistic 
content of the model, which is related to the strength of
$\,V\,-\,S\,$, is responsible for the shift of the Coester line to the
empirical region of nuclear matter saturation.

\end{abstract}


\maketitle


\centerline {\bf{1. INTRODUCTION}}
\bigskip

Quantum Chromodynamics (QCD) is the fundamental theory of the strong
interaction and hence it should be able to explain possible modifications of
hadron properties in the nuclear medium. However, typical nuclear
phenomena at intermediate and low energies cannot be analitically derived
from QCD, although one hopes that QCD will be solved numerically on the
lattice in the near future. Meanwhile, we are left with the construction
of phenomenological models in order to describe nuclear phenomena
and bulk properties. On these grounds, we can conceive models
in which hadrons are the degrees of freedom of some proposed Lagrangian
density. From the Lagrangian, a two-nucleon interaction can be obtained and
the parameters of this interaction are adjusted to reproduce the 
two-nucleon bound and scattering states observables. With this two-nucleon
interaction, the many-body problem can be solved via a Brueckner-Hartree-Fock 
(BHF) calculation  or a Relativistic-Brueckner-Hartree-Fock
(RBHF) \cite{bhf1,bhf2,bhf3}, if  one intends to incorporate relativistic
effects. 

Another phenomenological way to treat the many-body problem has been proposed 
by Walecka and collaborators \cite{wa,sw}. 
Their approach is based on a simple renormalizable field-theoretical model
which is often referred to as Quantum Hadrodynamics (QHD). In this model
nucleons interact through the exchange of $\sigma$ and $\omega$ mesons,
with the $\sigma$ providing medium range attraction and the $\omega$
the short-range repulsion. This model is usually solved in the mean
field approximation (MFA), in which the meson fields are replaced by their
expectation values. The Walecka model has achieved important goals in the
description of hadronic matter such as, for example, the calculation
of some bulk properties of nuclear matter as well as some properties of
finite nuclei. An interesting result of the model was to show the
relativistic mechanism for nuclear matter saturation: it occurs at a
density ($\rho_o$) at which the scalar (S) and the vector (V) potentials
largely cancel each other out. A curious aspect of this model is that the
masses of the scalar and vector mesons ( $m_\sigma$ and $m_\omega$), and
the coupling constants ( $g_\sigma$ and 
$g_\omega$) are eliminated in the equation of state for infinite 
nuclear matter in favour of\, 
$C_{\sigma}^2 = g_{\sigma}^2M^2/m_{\sigma}^2$\,  and
\, $C_{\omega}^2 = g_{\omega}^2M^2/m_{\omega}^2$ . 
These dimensionless constants, $C_{\sigma}^2$ and  $C_{\omega}^2$, the
only free parameters in this simple version,   
are fitted to reproduce the nuclear matter bulk properties.  
Besides the relativistic mean field calculation, the
Walecka model has also been used in a more complete treatment, the relativistic
Hartree-Fock approximation \cite{sw,rhf,fra}. 
A more sophisticated version of the Walecka model that includes
non-linear self-interactions of the scalar field has been
proposed by Boguta and Bodmer\cite{BO}. This non-linear version of the
Walecka model has obtained not only more reasonable results for the
incompressibility and
the effective nucleon mass  at nuclear matter saturation density  
but also has described various aspects of nuclear structure properties
\cite{RH,RH1}. This model with a very limited number of parameters
is also able to describe deformed nuclei \cite{RING,RING1} and for the 
first time  the anomalous shifts in the isotopic chains of different 
nuclei has been explained \cite{SHARMA,RING2}. As a consequence of all this
work, a new parametrization for this non-linear version of the Walecka
model has been proposed and gives a very good description not
only for the properties of stable nuclei but also for the nuclei far from
the  valley of beta stability \cite{RING3}.

The RBHF and Walecka models treat the many-body problem in  different 
perspectives.  The RBHF approach starts from a two-nucleon interaction 
which, in principle, fits some of the two-nucleon  observables and 
proceeds iterating an effective two-nucleon t-matrix up to the 
point where convergence is obtained to the nuclear matter saturation.
Therefore, depending on the two-nucleon interaction used, the nuclear
matter binding energy and the saturation density change. On the other 
hand, the Walecka model and some of its variants try first of all to fit 
directly the observed nuclear matter binding energy at the experimental
density. The tracing back of the two-nucleon interaction becomes difficult in 
this mean field scenario. To understand why, let's point out that the
two-nucleon interaction that emerges from the Walecka model 
in the static approximation is the sum of two Yukawa potentials, 
one attractive and the other repulsive. The
range of these potentials should be controlled by the masses of the
mesons, $m_{\sigma}$ and $m_{\omega}$. However, the mean field approach in
a Hartree perspective shows a dependence only on the dimensionless
constants $C_{\sigma}^2$ and $C_{\omega}^2$, as already pointed out, 
and we loose track of the value of $m_{\sigma}$ and $m_{\omega}$. Let us
recall that, in the mean field picture, this happens in a medium where
each nucleon feels equally an averaged constant interaction coming from all
the other nucleons. If one intends to have a crude idea about how a mean
field many-body calculation such as the Walecka model may be connected to
a two-body system in the vacuum, one has to model a two-body interaction
in a relativistic perspective. 
The simplest interaction should be the sum of two large potentials, a
scalar one $\,S\,$ and the time component of a vector one
$\,V\,$, each of them with an absolute value of a few hundreds MeV. 
In infinite nuclear matter, in the framework of the Walecka model, these
two components cancel each other out to a large extent in the calculation of 
the binding energy.  The assumption is that the same mechanism applies in
the two-nucleon system to calculate the two-nucleon binding energy. 
  
Calculations done for the nonrelativistic nuclear matter BHF model 
\cite{bhf1}, 
performed with different two-nucleon interactions, all of them fitting the 
deuteron binding energy and scattering data, show a correlation between 
the binding energy and
the corresponding saturation density. This correlation, known as
the Coester line, shows that the results obtained using different
nucleon-nucleon interactions predict nuclear matter saturation to be
located along a line which does not include the empirical data 
\cite{coester}.
Recent work \cite{rec} based on a relativistic model of the 
nucleon-nucleon interaction addresses the discussion of whether this same
correlation also takes place in a relativistic 
many-body calculation, exhibiting thus a kind of universality. We focus on the 
same question and, 
through a simple model, find that such a 
correlation is indeed present when the quantity $\,V\,-\,S\,$ is approximately 
constant. 
This last assumption can be justified since results from different hadronic 
models suggest that, 
in order to reproduce the experimental spin-orbit splittings for finite nuclei, 
 the $\,V\,-\,S\,$ quantity should lie in a narrow band \cite{furn3}. 
Moreover we conclude that the relativity within the model, which is related
to the strength of $\,V\,-\,S\,$, is responsible for the shift of the
Coester line to the empirical region of nuclear matter saturation. We see
this as an indication of a Dirac structure for the N-body potential with
large scalar and vector parts.

After having finished this work we became aware of the appearance of a
very recent paper 
analyzing this same correlation with the Bonn potentials; these involve
different types of meson exchanges, in a relativistic Brueckner
approach \cite{GFF}. Representing the in-medium on-shell T matrix
covariantly by five Lorentz invariant amplitudes, this last paper also
arrives at
the same conclusion as ours, namely that the new Coester lines are shifted
towards
the
empirical region of saturation supporting the Dirac structure of the
Nucleus-Nucleus potential.

Thus we can see that, despite dealing with a simple two-body model
that has only central terms and models nuclear matter with
the simple linear version of the Walecka model, we have the right physics,
which is embeded in the relativistic form of the potential with large 
scalar and vector parts.

\bigskip

\centerline {\bf {2. THE TWO-BODY MODEL}}
\bigskip
We start out with a two-body interaction consisting of a repulsive part
 $( \,V\, )$ and
an attractive one $( \,S\, )$. For simplicity, we will take the
potentials as constant inside a determined radius $\,R_o\,$ and zero
otherwise. The relativistic interaction is given by the operator 
$\,U\,=\,U_R\,=S\,+\,\gamma^{o}V\,,$ 
where $\,\gamma^{o}\,$ is the time component of the Dirac matrix and
$\,S\,$ is implicitly multiplied by the identity matrix. 
We simulate our two-body system 
by a one-body Dirac equation in spinorial form, 
\begin{equation}  
(\,E\gamma^{o}\,-\, {\mathbf \gamma \cdot p} \,
-  M  \, -\, U\,)\,\psi \,=\,0. 
\label{de}
\end{equation}
The above equation can be solved for $\,U\,=\,\gamma^o \Sigma\,$ 
( the relativistic case where $\,\Sigma\,$ is considered as the  
time component of a pure attractive vector interaction, hereafter 
labeled R1 ) or for $\,U\,=\,U_{R}\,$  
( the relativistic case with scalar and vector interactions,  R2 ). 
We have solved Eq. (\ref{de}) for each of these two potentials
 and for zero orbital angular 
momentum in coordinate space for different values of $\,R_o\,$, 
varying from 1.35 to 1.7 fm, obtaining different values for 
the two-body binding energy $\,B_{2}\, =\, E\, -\, M\,.$
Since we will later establish a connection with a N-body relativistic 
Walecka model, we have already imposed, for the case $\,U=U_R\,$, 
a constraint on $\,S\,$ and $\,V\,$ to satisfy the optical relationship 
required from the Hugenholtz-van Hove theorem for any relativistic
nuclear matter model in the mean field approximation \cite{hvh}, 
\begin{equation}
 - B_N\,=\, V\, +\, \left[k_{f}^2 \,+\,(M+S)^2 \,\right]^{1/2} \, - \,M, 
\label{vh}  
\end{equation}
where $\,B_{N}\,$ is the nuclear matter binding energy per nucleon, 
 $\,k_{f}\,$ 
the Fermi momentum at saturation density, and $\,M\,$ 
the bare nucleon mass. 
We have identified in the relativistic model R2 the quantity 
 $\,S+V\,$ with the depth of the square well vector potential
$\,\Sigma\,$. This identification comes often when one
interprets the Walecka model as qualitatively performing a 
large cancellation between $\,S\,$ and $\,V\,,$ resulting in a
value close to $\,-70 \,$ MeV for $\,\Sigma\,$. 
In Fig. (1) we display our results for the case
$\,B_{N}\,=\,$ 16 MeV,  $\,k_{f}\,= 1.3 fm^{-1} $, 
$\,R_{o}\,$= 1.4 fm, and  $\,M\,$= 939 MeV where,  
for the sake of comparison,  
a nonrelativistic Schroedinger equation calculation for the 
nonrelativistic interaction $\,U_{NR} \,=\, \Sigma\,$ is included.
 The curves show 
that the solution of the relativistic Dirac equation with the time 
component of a pure vector interaction changes the nonrelativistic Schroedinger result only 
slightly. The binding energy in
the former has a value somewhat greater than in the latter. However, the
Dirac equation solved with the relativistic interaction, containing scalar
and vector parts, changes dramatically the two-body binding in the 
opposite direction, i.e. the system becomes less bound. 

The results can be understood as follows. 
The three distinct cases can be presented through 
formal dispersion relation equations in terms 
of the large component $\chi$ of the wave function,
\begin{eqnarray}
  p^{2} \chi \,&=&\, 2\,M\,( e  \,-\, \Sigma\,)\, \chi,
~~~~~~~~~~~~~~~~~~~~~~~~~~~~~~~~~~\rm NR ~Model \label{nr} \\
 p^{2} \chi \,&=&\, 2\,M\,(\,e  \,-\, \Sigma\,)\,
(\,1\,+\, \frac{e -\Sigma}{2M}\,)\,\chi,~~~~~~~~~~~~~~~~ \rm R1 ~Model
\label{rl}
\\
 p^{2} \chi \,&=&\, 2\,M\,(e  \,-\, \Sigma\,)\,
(\,1\,+\, \frac{e -\Delta}{2M}\,)\,\chi,~~~~~~~~~~~~~~~~ \rm R2 ~Model
\label{sv}
\end{eqnarray}
where in the R2 model $\, \Delta\,=\,V\,-\,S\,$, $\, \Sigma = S+V \,$, and
in Eqs. (\ref{rl} - \ref{sv}) $\,e\,$ stands for the difference
between relativistic energy and the nucleon rest mass.

All these cases can be cast into an unified formula in terms of a 
nucleon effective mass $M^{*}$,
\begin{equation}
\frac{ p^{2}}{2M^{*}} \chi \,=\,( e  \,-\, \Sigma\,)\, \chi,
\label{eff} \\
\end{equation}
 where
\begin{eqnarray}
 M^{*} \,&=&\,\,M, ~~~~~~~~~~~~~~~~~~~~~~~~~~~~~~~~~~~~ \rm NR ~Model \label{m1} \\
 M^{*} \,&=&\,\,M\,(\,1\,+\, \frac{e -\Sigma}{2M}\,), ~~~~~~~~~~~~~~~~~~ \rm R1 ~Model \label{m2} \\   
 M^{*} \,&=&\, \,M\,(\,1\,+\, \frac{e -\Delta}{2M}\,). 
~~~~~~~~~~~~~~~~~~ \rm R2 ~Model \label{m3}
\end{eqnarray}
Since $\, \Sigma \,$ and $\, \Delta \,$ are negative and positive
quantities respectively, Eqs. (\ref{m1} - \ref{m3}) suggest
that the model R1 (where the effective mass is larger than $\,M\,$, 
leading to a smaller kinetic energy) should in fact lead to a more tightly bound system, 
with greater binding energy than model NR, whereas model R2 (where the effective mass is 
smaller than $\,M\,$, leading to a larger kinetic energy) 
is the case where the system is less bound. 
The same information could be inferred if one thinks in terms of the
 effective quantities  $   \Sigma ^{*} \,=\, \Sigma ( 1 - \Sigma/2M ) $
and  $\,e^{*}\,= \,e\,(\,1-\Sigma/M) $ for Eq.(\ref{rl}),
and  $\, \Sigma ^{*} \,=\, \Sigma ( 1 - \Delta/2M )$ and
$\,e^{*}\,= \,e\,(\,1-\Delta / M) $ for Eq.(\ref{sv}).
 In the cases we are discussing, the binding energy is small
 compared to the nucleon rest mass and we have neglected terms of order
$\,e^{2}/2M \,$. 
Note that for the model R2, not only $\, \Sigma \,$ 
 but also  $\, \Delta \,$, appear in Eq. ({\ref{m3}).
 The potential depth $\, \Sigma \,$ and the radius 
 $\, R_{o} \,$ control the two-body binding energy in Eqs. (\ref{m1})
  and (\ref{m2}), whereas in Eq. (\ref{m3}) an additional value 
  - $\, \Delta \,$ - is needed, working therefore as a new degree of 
 freedom of the problem, since for the same $\,R_{o}\,$, 
 different possible values of $\, \Sigma \,$ and  
 $\, \Delta \,$ are possible, as shown in Fig. 2. 
 The large difference between the results of the models R1 and R2 
 is not surprising, since $\, \Delta \,$ is expected to be 
 about ten times larger than $\, \Sigma \,$. 
To stress this point, it is worth mentioning that  
 the two-body bound state for the R1 model  
 arises when $\, - \Sigma \, R_{o}^2 \ge \pi^2\hbar^2 / 4M $  
 while for the R2 model the condition reads 
 $\, - \Sigma \,\Delta \, R_{o}^2 \ge \pi^2 \hbar^2 / 8\,$, 
 in units where $\, c = 1\ $. 

\bigskip

\centerline {\bf {3. THE N-BODY MODEL}}
\bigskip
The N-body model we use for nuclear matter is the linear Walecka model that 
now we briefly present.  
The degrees of freedom are baryon fields ($\psi$), scalar meson fields
($\sigma$), and vector meson fields ($\omega$). The Lagrangian density is
given by
\begin{eqnarray}
 {\cal L} &=& \bar \psi i \gamma_{\mu} \partial^{\mu } \psi -
\bar \psi M \psi + \frac{1}{2}(\partial_{\mu }\sigma
\partial^{\mu}\sigma - m^2_{s} \sigma^2) \nonumber \\
          &+& g_{s}\sigma \bar\psi \psi - 
 - \frac{1}{4}F^{\mu \nu}F_{\mu \nu} + \frac{1}{2} m^2_{v}\omega_{\mu}
\omega^{\mu} - g_{v}\bar \psi \gamma_{\mu }\psi \omega^{\mu },
\end{eqnarray} 
where $F_{\mu \nu} = \partial_{\mu}\omega_{\nu} - \partial_{\nu}\omega_{\mu} 
\,$, $\,M\,$ is the bare nucleon mass, and $\,m_{s}\,$ and $\,m_{v}\,$
are the scalar and vector mesonic masses respectively.
From the above Lagrangian we obtain, 
through the Euler-Lagrange formalism,
the equations of motion for the nucleon and mesons fields.

When the meson fields are replaced by the constant classical fields
$\sigma_{o}$ and $\omega_{o}$, we arrive at the mean-field approximation
with the equations
\begin{eqnarray}
  \omega_{o} &=& \frac{g_{v}}{m_{v}^2}\langle {\psi^+}{\psi}\rangle = 
\frac{g_{v}}{m_{v}^2}{\rho_{b}}, \label{og} \\
  \sigma_{o} &=& \frac{g_{s}}{m_{\sigma}^2} \langle \bar\psi\psi \rangle
                  =\frac{g_{s}}{m_{\sigma}^2}{\rho_{s}}, \label{sg} 
\end{eqnarray}
where $\rho_b$ and $\rho_s$ are the vector and scalar densities 
respectively.

   Now we define the scalar ($\,S\,$) and the vector ($\,V\,$)
potentials. This can be done by looking at the Dirac equation for
the models, and rewriting $M^{*}$ in the form 
$\, M^{*}\,=\,M -  g_{s}\sigma\,\,=\,M\,+\,S\,$.
Still from the analysis of the Dirac
equation, $V$ can be defined as a quantity which shifts the
energy, $\, V\,=\,g_{v}\omega_{o}\,$.  
It is convenient to introduce the following dimensionless quantities:
 $\, m^{*} \,=\, M^{*}\,/\,M \,\,$,
 $\,\,C_{s}^2 = g_{s}^2M^{2}/m_{s}^2\,\,$,  and
 $C_{v}^2 = g_{v}^2M^{2}/m_{v}^2\,\,.$
Using Eqs. (\ref{og}) and (\ref{sg}) we can rewrite the 
scalar and vector potentials in
terms of $C^2_s$ and $C^2_v$ as 
\begin{equation}
S= -\,\frac{C^2_s}{M^2}\rho_s \,, \hspace{2cm}
\,V=\frac{C^2_v}{M^2}\rho_b \,.
\end{equation}

The expressions for the energy density and pressure at zero
temperature can be found as usual by the average of the
energy-momentum tensor,
\begin{equation}
{\cal E} = \frac{C_{\omega}^{2}}{2M^{2}}\rho^2+
\frac{M^{4}}{2C_{\sigma}^{2}}(1-m^{*})^2+
\frac{\gamma}{2\pi^2}\int_{0}^{k_f} k^2dk\,E^{*}(k),
\end{equation}
and
\begin{equation}
  p=\frac{C_{\omega}^{2}}{2M^{2}}\rho^2-
\frac{M^{4}}{2C_{\sigma}^{2}}(1-m^{*})^2+
\frac{1}{3}\frac{\gamma}{2\pi^2}\int_{0}^{k_f} k^2dk\,
\frac{k^{2}}{E^{*}(k)},
\end{equation}
where $\, \rho \, = \,(\gamma/6\pi^2)k_{f}^3\,$,
 $ \gamma = 4$ is the degeneracy factor for symmetric nuclear matter,
and $E^{*}(k)\,=\,( k^2 + M^{*^2} )^{1/2}\,$.

The solution of the model is obtained explicitly
through the minimization of ${\cal E }$ as a function of $m^{*}$. The
equation thus obtained reads
\begin{equation}
 1-m^{*}-\frac{\gamma}{2\pi^2}\int_{0}^{x_f}\,\frac{x^2dx}
{(x^2+m^{*^2})^{\frac{1}{2}}}\,\,=0\,\,,\,\
\label{gap}
\end{equation}
where we have introduced the dimensionless variable $ x=\frac{k}{M}$.
This equation has to be solved self consistently and provides the
basis for obtaining all kinds of thermodynamical quantities in
the mean field approach we are using.

The usual procedure to obtain the values of the coupling constants 
is the following: at the saturation density $\,\rho_o\,$ the pressure vanishes 
( hydrostatic equilibrium )
and at this same point the Hugenholtz-van Hove theorem allows us to write
down the relation, already built-in in Eq.(2), 
 ${ \cal E} / \rho_o\, = \,V\, + \,\left[k_{f}^2 \,+\,(M+S)^2\right]^{1/2} 
  - \, M.$ These two constraints can then be used to find the 
 constants $C_{\sigma}^2$ and  $C_{\omega}^2$ by  imposing the 
 experimental values for the nuclear matter binding energy 
$\,{\cal E} / \rho_o - M = -B_{N} $ = -16 MeV and
for the density $\, \rho_o \,=\,0.15 fm^{-3}\,$. 

We could have used the non-linear version of the   
Walecka model discussed in the introduction \cite{RING3}.
This would have changed the expression for the energy density and the self
consistent Eq.~(\ref{gap}); however, the correlation we are analyzing in 
this work would still survive, because of the presence of the large scalar
and vector potentials.

\bigskip

\centerline {\bf {4. RESULTS AND DISCUSSIONS}}
\bigskip
To look for the correlation between two-body and N-body
bindings we proceed in the following way. From  the
two-body calculation for a fixed two-body binding energy $B_2$ at a given 
radius $\,R_{o}\,$  one obtains the scalar ($\,S\,$)  
and the vector ($\,V\,$) potentials. 
In the Walecka model we then use these two quantities, 
coming from the two-body calculation, in order to find the constants
$C_{\sigma}^2$ and  $C_{\omega}^2.$  Only after we have these constants
 do we proceed in the calculation of the N-body system saturation.
Therefore, we will have a set of saturated ($\,B_{N}\,,\, \rho_o \,$)  
pairs for each different ($\,S\,, \,V\,$) set pairs,   
 which  keeps a fixed $B_2$. We have chosen a fixed 
 $\,B_{2} \,$ = 2 MeV, allowing variation of $\,R_{o}\,$ from 
 1.4 fm to 2.1 fm. The set of $\,(R_{o},\,S\,,V\,$) is given by 
 Fig. 2 in a three-dimensional plot. 
 
  Within the above procedure we have obtained the set of 
  ($\,B_{N}\,,\rho_{o}\,$) values presented in Fig. 3. In principle, 
  there is no clear correlation between $\,B_{N}\,$ and 
  $\,\rho_{o}\,$ unless one constrains the interaction 
  of the two-body system to have the same range $\,R_{o}\,$ but 
  allowing different values of $\,\Delta\,$, or to have the same 
  value of $\,\Delta\,$ but having different $\,R_{o}\,$ values. In
  the first case, the correlation points show a positive slope for
  increasing saturation density while in the second one, the Coester 
  line correlation is present. This correlation has been shown to exist 
   in nonrelativistic calculations of the nuclear matter saturation 
   point. Different phenomenological two-body potentials furnish different  
   ($\,B_{N}\,,\rho_{o}\,$) values, which lead to a rough
   linear increasing of $\,B_{N}\,$ for increasing $\,\rho_{o}\,$ also shown 
   in the Fig.3. 
   To trace back the reason why this happens in the relativistic case, we 
   have rewritten Eq. (\ref{vh}) as 
\begin{equation}
 - B_{N}\,=\,\frac{k_{f}^{2}}{2M^{*}} \,  + \, \Sigma \,,
\label{vh2}  
\end{equation}   
 where   $\,M^{*}\,=\,M\,(1\,-\,\frac{B_{N}+\Delta}{2M})$. 
  Since $B_{N} \ll \Delta \,$, $\,M^{*}\,  \rightarrow  
   M\,(1\,-\,\Delta/2M)$ and when $\,\Delta\,$ becomes 
   constant, Eq. (\ref{vh2}) acquires a nonrelativistic 
   form with an effective nucleon mass. The relativistic information of the
model is contained in the strenght of $\,\Delta\,$. The relativistic
effect is shown in Fig. 3, where the Coester line is
shifted to the empirical region of saturation provided $\,\Delta\,$ is
large. This suggests a Dirac structure for the potential with large scalar
and vector parts. 

  Regarding the role of $\,\Delta\,$ in the nuclear data fittings, 
  it is interesting to point out that recent nuclear matter analysis with 
 many different quantum-hadrodynamics models show a correlation 
between the finite nuclei spin-orbit energy splitting and 
the nuclear matter effective nucleon mass $\,m^{*}\,$ \cite{furn3}. 
The spin-orbit energy splitting increases as $\,m^{*}\,$ decreases. 
To accurately reproduce the empirical splittings, it is found that  
one should require $\,m^{*}\ $ to be between 0.58 and 0.64. 
From the Hugenholtz-van Hove theorem \cite{hvh}, we can 
relate $\,m^{*}\,$ and $\,\Delta\,$ through  
$\,\Delta\,=\,2M\,-B_{N}\,-M^{*}\,-\,( k_{f}^2 + M^{*^2} )^{1/2}\,$.
Therefore, a correlation between the splittings and 
$\,\Delta\,$ also exists, namely the energy splitting increases with 
$\,\Delta\,$.  Moreover, the phenomenology of finite nuclei restricts  
 the values of $\,\Delta\,$ to lie in a narrow band 
 if one intends to reproduce good results for spin-orbit splitting. 
  On the other hand, 
  relativistic results, obtained via Dirac-Brueckner-Hartree-Fock 
   calculations, using three different versions (A,B and C) of the 
   Bonn potential, show a roughly linear $\,B_{N}\, \times \rho_{o}\,$ 
   correlation \cite{bhf1}.  The corresponding nuclear matter 
   $\,\Delta\,$ quantities lie on a narrow range.    

In Fig. 4 we present the incompressibility ($\,K\,$) as a function of 
$\,\rho_{o}\,$. Again, the possible correlations follow the same trend  
established in Fig. (3).  Roughly, $\,K\,$ increases linearly 
with $\,\rho_{o}\,$ for fixed $\,\Delta\,$ values.  
When we consider bigger values for $\,\Delta\,$, $\,K\,$ increases, turning the 
equation of state stiffer. 
In Fig. (5), the effective nucleon mass $\,M^{*}\,$ is displayed 
as a function of $\,\rho_{o}\,$, also showing the same qualitative 
behavior for the possible correlations in terms of the 
two-body $\,R_{o}\,$ and $\,\Delta\,$ inputs. As expected by our previous
discussion, when $\,\Delta\,$ increases $\,M^*\,$ decreases.
In these two last figures we only want to exhibit the correlations between
the incompressibility and the effective nucleon mass on one hand and the
saturation density on the other. The specific values of the two first
quantities are not important here and we would certainly get more
reasonable results for them at the nuclear matter saturation density had
we used the non-linear Walecka model\cite{RING3}.

\bigskip

\centerline {\bf {5. CONCLUSIONS}}  
\bigskip
 The plots of the binding energy versus the saturation density 
 of nuclear matter that are obtained by nonrelativistic calculations 
 with a large number of different two-nucleon interactions 
 show a correlation that is known as the Coester line. 
 In a nonrelativistic approach, the Coester line has to be conceived  
from the phenomenology contained in two-body systems with a low 
fixed binding energy interacting via a short range potential. 
In terms of our simple two-body model, the main ingredients 
are  $\Sigma$ ( the depth of the 
potential) and the range $R_{o}$. Both suffice to parametrize
in a nonrelativistic approach
  the two-body problem itself, with a fixed two-body binding energy.  

From a relativistic point of view, we have presented a discussion about 
this possible correlation.  Our main conclusions can be summarized 
as follows: 

(1) In the relativistic case with scalar and vector interactions, 
 a new degree of freedom $\,\Delta \,$ 
 arises and we have seen that unless $\,\Delta\,$ is constant, the 
 correlation contained in the nonrelativistic Coester line is destroyed. 
 
 (2) Since the phenomenology of relativistic finite nuclei 
 calculations restricts  
 the values of $\,\Delta\,$ to lie in a narrow band,  
 if one intends to reproduce good results for spin-orbit splitting, 
  we conjecture that a Coester line would survive in such models. 

  (3) For large values of $\,\Delta \,$ we show that the relativistic 
  Coester line is shifted to the empirical region of saturation, which we
  see as an indication of a Dirac structure for the potential with large
  scalar and vector parts, and also as a signature of the relativistic
  content of nuclear matter.

  (4)  For a fixed value of $\,B_{2}\,$,  
  the study of $\,B_{N}\,,K\,$ and $\,M^{*}\,$ as a function
  of $\,\rho_{o}\,$ suggests that they are themselves correlated. 
  
  We have modeled our system in a very
 simple way, where the two-body interaction has only the central term.
 We are aware that the phenomenological two-nucleon interactions 
 in vacuum, as used by Reid, Paris, Hamada-Jonston, Bonn and others  
 have not only simple central terms, but include 
spin-spin, spin-orbit and tensor force. Their magnitudes are different in
strength. In particular, the tensor force is believed to be important
for the nuclear matter saturation \cite{tjon}. 
However, as already mentioned in our introduction, a recent paper has
analyzed the Coester line for various Bonn potentials in a
relativistic Brueckner approach \cite{GFF}. Its conclusion is that the
new Coester lines are shifted towards the empirical region of saturation,
supporting a Dirac structure for the Nucleus-Nucleus potential.
Then we can conclude that even when we are dealing with a simple two-body
model that has only central terms, we have the dominant aspect of
the correct physics. This physics is in fact embedded in the relativistic
form of the potential, with scalar and vector parts, because, in the
medium, many pieces of the interaction are averaged out, while the
remaining part is largely dominated by the scalar and the vector central
potentials \cite{bernar}.

To finish, we remark that for the first time a connection between the
Coster line has been proposed as a function of the $\Delta = V\,-\,S$
quantity, Fig. 3. We have exagerated the $\Delta$ interval in this figure
to better exhibit how the relativistic content of the model would manifest
itself in the Coester lines themselves.

\section*{ACKNOWLEDGEMENTS}
We would like to thank Dr. Yuki Nogami for the careful reading of our 
manuscript and his very useful sugestions.

\clearpage

\begin{figure}[t]
\centering
\includegraphics[width=0.6\textwidth , height=0.4\textheight]{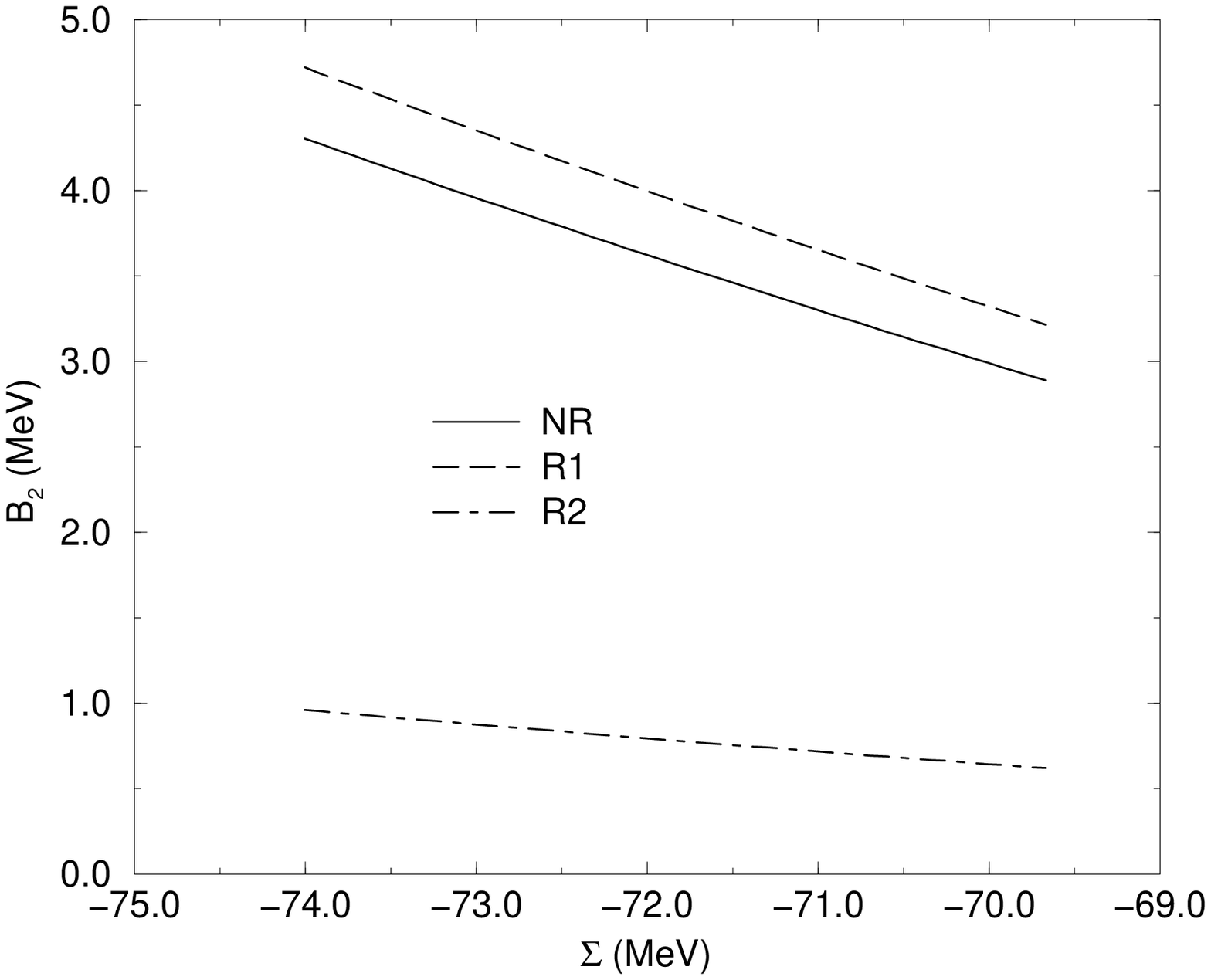} 
\caption{Two-body binding energy $\,B_{2}$ as a function of
$\Sigma = V+S\,$ for different approaches, 
Nonrelativistic ($NR$), Relativistic Dirac with pure vector 
interaction ($R1$) and Relativistic Dirac with scalar and vector
interaction ($R2$).}
\label{fig1}
\end{figure}

\begin{figure}[b]
\centering
\includegraphics[width=0.6\textwidth , height=0.4\textheight]{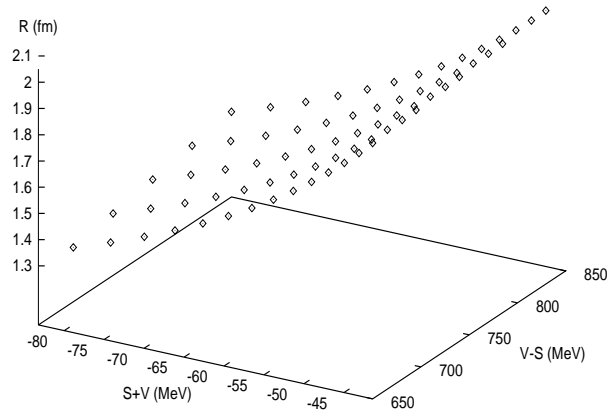} 
\caption{Set of $R=R_o$, $S+V=\Sigma$, and 
$V-S=\Delta$ values, with the constraint $B_2 = 2 \; MeV$, from the R2 model.}
\label{fig2}
\end{figure}

\begin{figure}[t]
\centering
\includegraphics[width=0.6\textwidth , height=0.4\textheight]{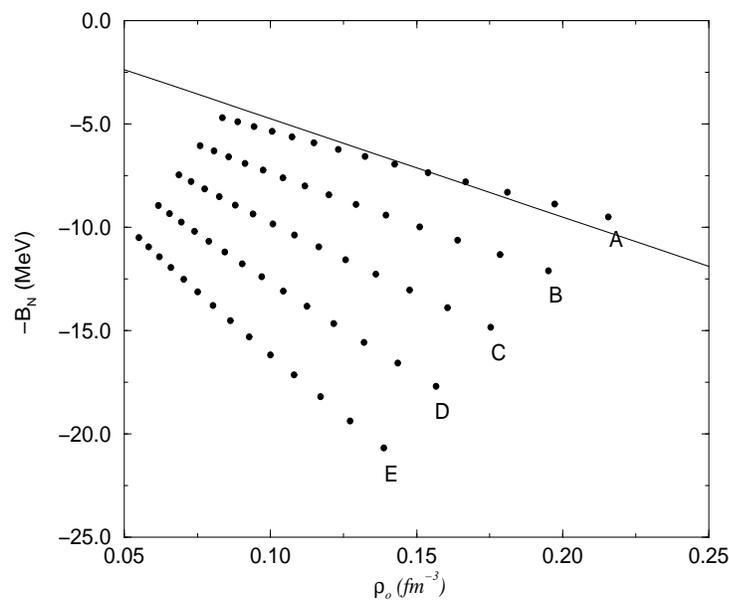} 
\caption{N-body binding energy $\,B_{N}$ as a function of
the density $\rho_o$ for different values of $\Delta$. 
The curves correspond to $\Delta$ = 650 (A), 700 (B), 750 (C), 800 (D), 
and 850 (E)  in MeV. The points on each line, from left to right, 
correspond to values of $R_o$ ranging from 1.4 to 2.1, in steps of
0.05 fm. All the points correspond to ($\,R_{o},\,S\,,V\,$) values 
leading to $\,B_{2}\,$= 2 MeV. The straight line represents the 
fitting of the nonrelativistic Coester line of ref. 1.} 
\label{fig3}
\end{figure}

\begin{figure}[b]
\centering
\includegraphics[width=0.6\textwidth , height=0.4\textheight]{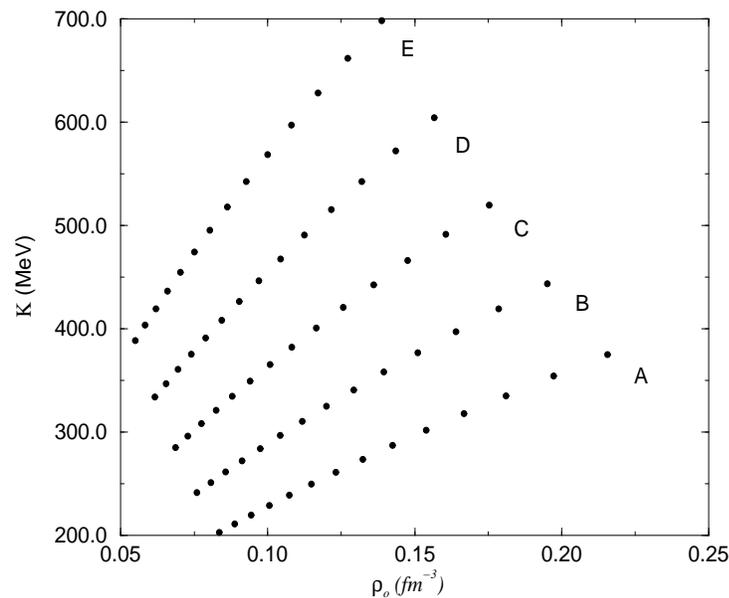} 
\caption{Incompressibility $K$ as function of the density
$\rho$. $\Delta$ and $R_o$ are the same as in Fig. 3.}
\label{fig4}
\end{figure}

\begin{figure}[t]
\centering
\includegraphics[width=0.6\textwidth , height=0.4\textheight]{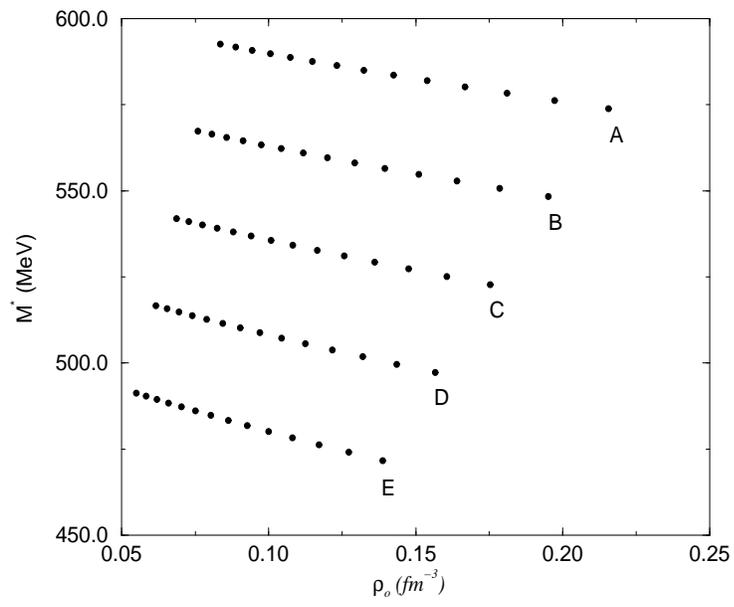} 
\caption{Nucleon effective mass as a function of
the density $\rho_o$. $\Delta$ and $R_o$ are the same as in Fig. 3.}
\label{fig5}
\end{figure}

\end{document}